*Title Page*

# Mid-infrared computational temporal ghost imaging

Han Wu,[1,†] Bo Hu,[1,†] Lu Chen,[1] Fei Peng,[2,*] Zinan Wang,[3] Goëry Genty,[4,*] and Houkun Liang[1,*]

*1 College of Electronics and Information Engineering, Sichuan University, Chengdu, Sichuan 610064, China*

*2 College of Electrical Engineering, Sichuan University, Chengdu, Sichuan 610064, China*

*3 Key Lab of Optical Fiber Sensing & Communications, University of Electronic Science & Technology of China, Chengdu, Sichuan 611731, China*

*4 Laboratory of Photonics, Tampere University, FI-33014 Tampere, Finland*

*†Han Wu and Bo Hu contribute equally to this work.*

Han Wu, Email: hanwu@scu.edu.cn

Bo Hu, Email: hubo_uestc@hotmail.com

Lu Chen, Email: inkest@163.com

Fei Peng **(Corresponding author)**, Email; pengfei@scu.edu.cn

Zinan Wang, Email: znwang@uestc.edu.cn;

Goëry Genty **(Corresponding author)**, Email: goery.genty@tuni.fi;

Houkun Liang **(Corresponding author)**, Email: hkliang@scu.edu.cn

**Abstract**

Ghost imaging in the time domain allows for reconstructing fast temporal objects using a slow photodetector. The technique involves correlating random or pre-programmed probing temporal intensity patterns with the integrated signal measured after modulation by the temporal object. However, the implementation of temporal ghost imaging necessitates ultrafast detectors or modulators for measuring or pre-programming the probing intensity patterns, which are not available in all spectral regions especially in the mid-infrared range. Here, we demonstrate a frequency downconversion temporal ghost imaging scheme that enables to extend the operation regime to arbitrary wavelengths regions where fast modulators and detectors are not available. The approach modulates a signal with temporal intensity patterns in the near-infrared and transfers the patterns to an idler via difference-frequency generation in a nonlinear crystal at a wavelength where the temporal object can be retrieved. As a proof-of-concept, we demonstrate computational temporal ghost imaging in the mid-infrared with operating wavelength that can be tuned from 3.2 to 4.3 μm. The scheme is flexible and can be extended to other regimes. Our results introduce new possibilities for scan-free pump-probe imaging and the study of ultrafast dynamics in spectral regions where ultrafast modulation or detection is challenging such as the mid-infrared and THz regions.**

**Introduction**

Ghost imaging originally emerged in the spatial domain as a correlation technique for imaging objects. The image is obtained by correlating the intensity of a reference beam which does not interact with the object itself but has its recorded spatial intensity distribution with the intensity of a test beam that illuminates the object and is measured by a single-pixel detector. This approach can offer significant advantages, including e.g. robustness against noise and distortions and enhanced security[1-5]. The probing patterns used in the reference beam can either be random and captured with a high-resolution detector or pre-programmed for computational ghost imaging[6-9]. Single-pixel imaging has been demonstrated in various wavelength regimes, and very recently at the single-photon level in the mid-infrared region using frequency upconversion[10].

In the past few years, the concept of ghost imaging has expanded beyond the spatial domain, into the temporal[11] and spectral domains[12,13]. Temporal ghost imaging (TGI) utilizes random temporal intensity fluctuations from a chaotic light source such as e.g. pseudo-thermal light or multimode laser as a reference probe signal. This signal is then modulated by a temporal object, and the object is subsequently reconstructed from the correlation between the random intensity fluctuations and the integrated intensity after the temporal object measured by a slow detector. Enhanced TGI schemes have also been developed including differential TGI[14] and magnified TGI[15] to improve the signal-to-noise ratio and temporal resolution of TGI, or Fourier TGI[16] that can provide additional spectral information on the temporal object. TGI has generated considerable interest across various applications, including secure communication, underwater communication, and quantum device characterization[17-20]. Furthermore, similarly to computational ghost imaging in the spatial domain, computational TGI can be implemented by utilizing ultrafast modulators to pre-program temporal patterns rather than utilizing random temporal intensity fluctuations[21-23]. It is even possible to utilize the finite time-varying response of slow detectors to perform one-time readout TGI, enabling the use of very slow optoelectronic detection devices including light-emitting diodes or solar cells [24,25]. However, the absence of suitable instrumentation, such as ultrafast mid-infrared electro-optic modulators for pre-programming temporal patterns at mid-infrared light sources, has been a bottleneck in the direct implementation of computational TGI in the mid-infrared.

When compared to direct detection methods, TGI for retrieving ultrafast temporal objects offers several important advantages. One key benefit is that TGI does not require to precisely resolve the temporal profile of the probing signal after interacting with the object, making the technique intrinsically insensitive to temporal distortion effects that may occur during propagation after the temporal object[11]. Additionally, because the measurement is based on integrated intensity data that can be collected with low-bandwidth photodetector that typically exhibit high sensitivity, TGI enables the retrieval of temporal objects at substantially reduced optical power levels. This feature can prove to be particularly advantageous in scenarios characterized by significant transmission losses or light scattering for example.

The original demonstrations of TGI systems have operated in the near-infrared region, leveraging mature near-infrared low-coherence fiber lasers, telecom ultrafast modulators, and detectors[11,14,15,22,23,26]. This has hindered the exploitation of the full potential of TGI. To extend the wavelength regime of TGI, a two-color scheme that employs second-harmonic generation has been recently introduced[27]. In this approach, the probing intensity fluctuations generated from a multimode quasi-continuous-wave (CW) laser source are converted in a nonlinear crystal to the second harmonic wavelength where they can be measured in real-time with a fast detector. This technique, however, requires a laser source with random intensity fluctuations at the wavelength of the temporal object as well as a large number of measurement realizations. In principle, one could employ pre-programmed patterns instead of random intensity fluctuations, but this would require ultrafast intensity modulators at the wavelength of the temporal object, which may not be available. A particular region of importance where TGI has yet to be realized due to the lack of suitable detectors and fast modulators is the mid-infrared, and expanding the operational range of TGI to the mid-infrared region holds particular promise for applications in e.g. free-space communication[28] and ultrafast pump-probe experiments in areas such as all-optical modulation[29] and semiconductor carrier lifetime measurements.

In this paper, we fill this gap and introduce the concept of frequency downconversion TGI that enables the experimental realization of computational TGI in the mid-infrared. Specifically, instead of directly pre-programming temporal patterns at mid-infrared wavelengths, the approach modulates pre-programmed temporal patterns at near-infrared wavelengths using a conventional telecom modulator and, subsequently, these modulated patterns are transferred to a mid-infrared idler via difference-frequency generation (DFG) using a temporally stable continuous-wave (CW) near-infrared pump light source. The approach is generic and can be implemented to other wavelengths regions where fast modulators and/or detectors are not available. As a proof-of-concept demonstration, we implement computational TGI in the 3.2 to 4.3 μm range using only one mid-infrared slow detector to image a temporal object generated from the on/off keying transmission of an acousto-optic intensity modulator (AOM). We anticipate that the proposed concept of frequency downconversion-based ghost imaging can open up new possibilities for achieving ultrafast computational imaging at wavelengths where conventional pre-programmed modulation techniques are challenging to apply, particularly in the mid-infrared and terahertz regions[30-32].

**Results**

**Experimental Setup.** A schematic of the experimental setup is illustrated in **Fig. 1** (see also **Materials and methods** for details). Light from a CW diode operating at 1542 nm is amplified by a commercial erbium-doped fiber amplifier (EDFA) with a maximum power of 200 mW. The 1542 nm CW light is temporally modulated by a 1.5 μm AOM (AOM1), which is driven by the pre-programmed patterns generated from an arbitrary waveform generator (AWG). The modulated 1542 nm light and a 4 W home-built ytterbium-doped fiber laser[33] tuned at 1060 nm used as a strong pump are collimated with fiber collimators (FC), and the two collimated beams are spatially multiplexed by a dichroic mirror. The ytterbium-doped fiber laser layout and spectral characteristics are provided in **Supplementary note S1.** The overlapping beams are then collinearly focused with a lens of 75 mm focal length into a periodically-poled $LiNbO_3$ crystal (PPLN, HC Photonics) with antireflection coatings at pump, signal and idler wavelengths for DFG process that converts the modulated 1542 nm signal to the idler at 3.4 μm. The idler is then collimated by a $CaF_2$ lens, and an antireflection-coated Germanium filter is used to block the residual near-infrared beams. The output power of the idler is about 100 μW and the pre-programmed temporal patterns modulating the 1542 nm signal are effectively transferred to the idler light through the DFG process.

To demonstrate proof-of-concept mid-infrared computational TGI, the transmission of a mid-infrared AOM (AOM2) driven by a bit sequence produced by the AWG is used as the temporal object to be retrieved. The mid-infrared idler light transmitted through AOM2 is detected by a slow photodetector (Thorlabs, PDAVJ5, 2.7-5.0 μm, 1 MHz bandwidth maximum) that cannot resolve the high-speed bit sequence and recorded by a real-time oscilloscope (RS, RTO2024, 2 GHz bandwidth) triggered by the AWG. The temporal object is then recovered from the correlation operation between the intensity recorded by the slow detector and the pre-programmed patterns generated by the AWG.

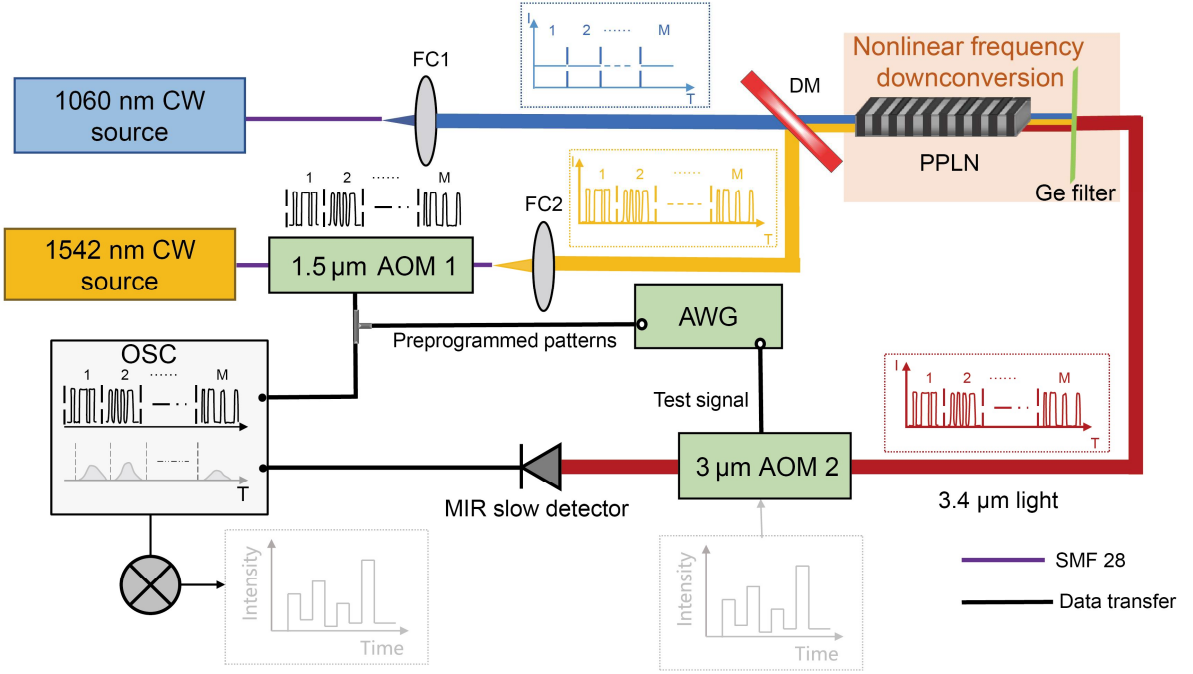

**Fig. 1** Experimental setup for computational temporal ghost imaging based on frequency downconversion. Preprogrammed probing temporal patterns generated from the arbitrary waveform generator (AWG) are applied to an acousto-optic modulator operating in the 1.5 μm range (AOM1) to modulate the 1542 nm laser. The probing patterns are transferred to an idler wavelength at 3.4 μm using downconversion in a periodically poled lithium niobate crystal (PPLN) with a strong 1060 nm CW pump. The 3.4 μm idler light is transmitted through the temporal object generated by acousto-optic modulator operating in the 3 μm range (AOM2) and detected by a mid-infrared slow detector (MIR slow detector). The temporal intensity from the slow detector and the preprogrammed temporal probing patterns are recorded by the oscilloscope (OSC) and the temporal object is retrieved from their correlation. FC, fiber collimator; DM, dichroic mirror (HR@1135–1600 nm, HT@1060 nm).

**Frequency downconversion of pre-programmed temporal patterns.** The core of mid-infrared computational TGI based on frequency downconversion is to transfer the preprogrammed temporal probing patterns from the near- to mid-infrared spectral region. When using DFG as the nonlinear conversion mechanism, the time-dependent intensity of the generated idler $I_{idler}(t)$ is proportional to $I_{idler}(t) \propto I_{pump}(t) \times I_{signal}(t)$ where $I_{pump}(t)$ and $I_{signal}(t)$ represent the time-dependent intensity of the strong pump at 1060 nm and signal at 1542 nm, respectively. Since the temporal intensity of the 1542 nm signal is modulated by the preprogrammed pattern and the intensity of the 1060 nm pump is continuous and stable over time, one expects that the temporal intensity profile of the idler follows that of the pre-programmed probing patterns modulating the signal beam. To confirm this, we first performed a direct comparison between the time-resolved intensity profile of the 1542 nm signal modulated by a sequence of pre-programmed but randomly selected binary patterns and that of the 3.4 μm idler after the PPLN crystal. Due to the limited bandwidth (1 MHz) of the mid-infrared photodetector used for the comparison, the speed of the AOM1 that modulates the 1542 nm signal was set to 625 kbps. Three examples of bit sequences recorded over a time window of 51.2 μs time at 1542 nm (red dash line) and the corresponding downconverted idler temporal profile at 3.4 μm (black solid line) are shown in **Fig. 2(a)**. One can see how the temporal modulation of the idler at 3.4 μm nearly perfectly follows that of the signal at 1542 nm. This near-perfect correspondence is further confirmed in **Fig. 2(b)** showing the time-to-time intensity cross-correlation map between the bit sequence modulating the 1542 nm signal and the recorded modulation at 3.4 μm calculated over 250 consecutive 0-1 random probing sequences. These results confirm that pre-programmed probing patterns can be successfully transferred to the idler at 3.4 μm by DFG, and that the idler carries all the necessary information to perform computational TGI in the mid-infrared.

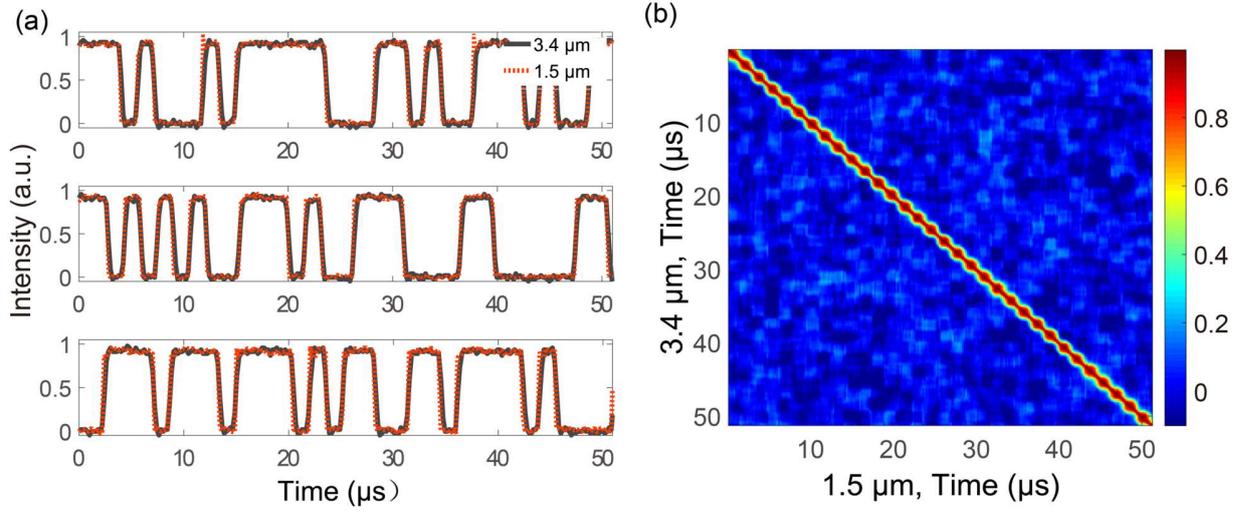

**Fig. 2** (a) Comparison of time-resolved intensity profiles of 1542 nm light modulated by the pre-programmed randomly selected binary patterns (red) and 3.4 μm idler temporal intensity after the PPLN crystal (black). (b) Time-to-time intensity fluctuations cross-correlation between the 1542 nm signal and 3.4 μm idler calculated over 250 temporal windows containing different probing patterns.

**Mid-infrared computational TGI with random patterns.** We first demonstrate computational TGI at 3.4 μm by using pre-programmed but randomly chosen binary patterns. The preprogrammed random patterns are illustrated in **Supplementary Note S2**. To obtain the ghost image of the temporal object, N distinct 0-1 bit sequences $R_k(t)$ ($k$ = 1, 2, …, N) generated by the AWG modulate the transmission of AOM1 and therefore the temporal intensity of the 1542 nm signal. The bit sequences that are randomly selected are then transferred to the idler at 3.4 μm via DFG in the PPLN crystal. The temporal object $S(t)$ with duration $T$ is also generated by the AWG and drives the transmission of AOM2 that modulates the idler intensity at 3.4 μm. Note that the temporal object is repeated periodically with a period equal to the temporal window of one probing bit sequence. The intensity integrated over a particular bit sequence $R_k(t)$ after AOM2 is recorded by the slow detector as $B_k$ and the temporal object is reconstructed by computing the second-order correlation between $R_k(t)$ and $B_k$ over the N distinct bit sequences as $S(t) = \langle R(t)B \rangle - \langle R(t) \rangle \langle B \rangle$ where $\langle \cdot \rangle$ denotes averaging over the $N$ probing patterns.

**Figure 3(a)** shows the experimental result obtained by correlating the binary probing patterns with the integrated intensity at 3.4 μm over 250 realizations for a temporal object with a modulation speed of 625 kbps. The modulation speed of the probing patterns was set to 1.25 Mbps, twice that of the temporal object. For comparison, we also performed a direct control measurement of the temporal object. For this purpose, the idler at 3.4 μm is generated by DFG without modulating the 1542 nm signal and directly detected with the 1 MHz bandwidth mid-infrared photodetector after modulation by the temporal object imposed by mid-infrared AOM2. The temporal object speed at 625 kbps was intentionally chosen such that the mid-infrared photodetector speed is sufficient to directly resolve the temporal intensity of the modulated 3.4 μm idler. One can see a very good correspondence between the ghost image (cyan solid line) and the direct detection (blue dashed line), confirming the ability of the downconversion TGI scheme to reconstruct temporal object in the mid-infrared regime. To illustrate the benefit of the downconversion TGI technique, we show in **Fig. 3(b)** the comparison of TGI and direct detection for a temporal object generated with a faster speed of 5 Mbps. The speed of the probing patterns in this case is set to 10 Mbps corresponding to a temporal resolution of 0.1μs. Due to the limited bandwidth of the mid-infrared detector, the structure of the temporal object cannot resolve by direct detection in this case as seen from the blue dashed line in **Fig. 3(b)**. Remarkably, however, we can see how the ghost image of the temporal object (cyan solid line) can resolve very well the structure of the temporal object. **Fig. 3(c)** and **(d)** shows additional examples of 5 Mbps temporal objects with different temporal structures resolved by TGI with again good agreement between the ghost images and the temporal objects.

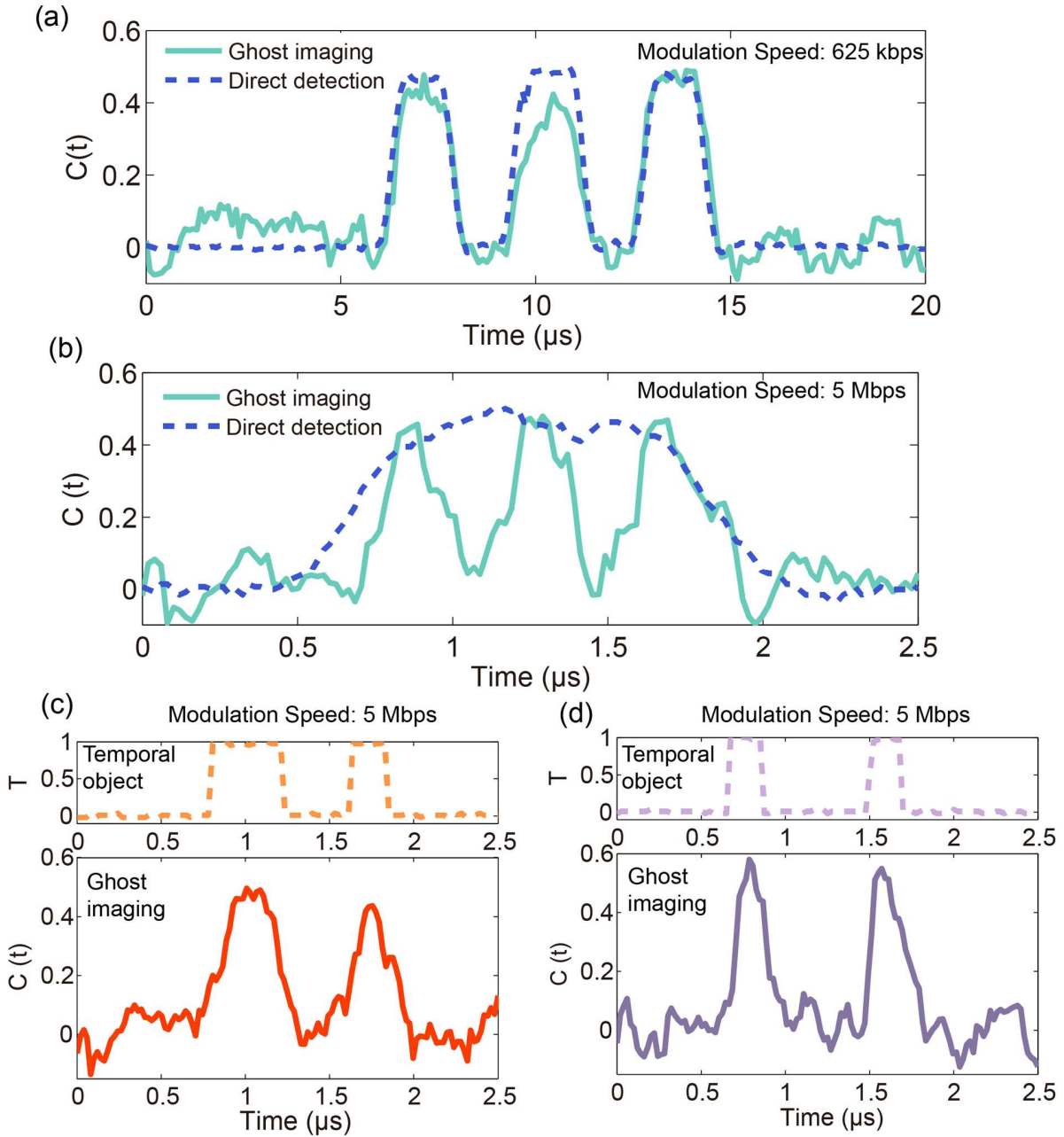

**Fig. 3** Mid-infrared TGI results using pre-programmed, randomly selected probing patterns. (a) Experimental ghost image (cyan solid line) of a temporal object with a modulation speed of 625 kbps retrieved from 250 distinct temporal probing patterns. (b) Experimental ghost image of a temporal object with a modulation speed of 5 Mbps. In (a) and (b) the blue dashed line corresponds to direct detection with a 1 MHz bandwidth mid-infrared photodetector. (c) and (d) shows additional TGI results of other examples of 5 Mbps temporal objects. The orange and purple dashed lines represent the corresponding bit sequences measured at the output of the AWG.

The quality of the retrieved ghost image as a function of the number of probing temporal patterns is investigated in **Fig. 4**. The speed of the temporal object is set to 625 kbps to allow for comparison of the retrieved ghost images (cyan solid lines) with direct detection (blue dashed lines). As can be expected, the signal-to-noise ratio of the TGI increases with the number of probing patterns, and it is found that the temporal object can be already resolved with about 160 distinct patterns. Improvement in the accuracy of the retrieved ghost image for an increasing number of realizations can be quantified by evaluating the peak signal-to-noise ratio (PSNR) between the directly measured and reconstructed temporal sequences[22] (see **Materials and methods** section). The PSNR values are 13.89 dB, 15.89 dB, 16.94 dB and 17.45 dB with 120, 160, 200 and 250 realizations, respectively.

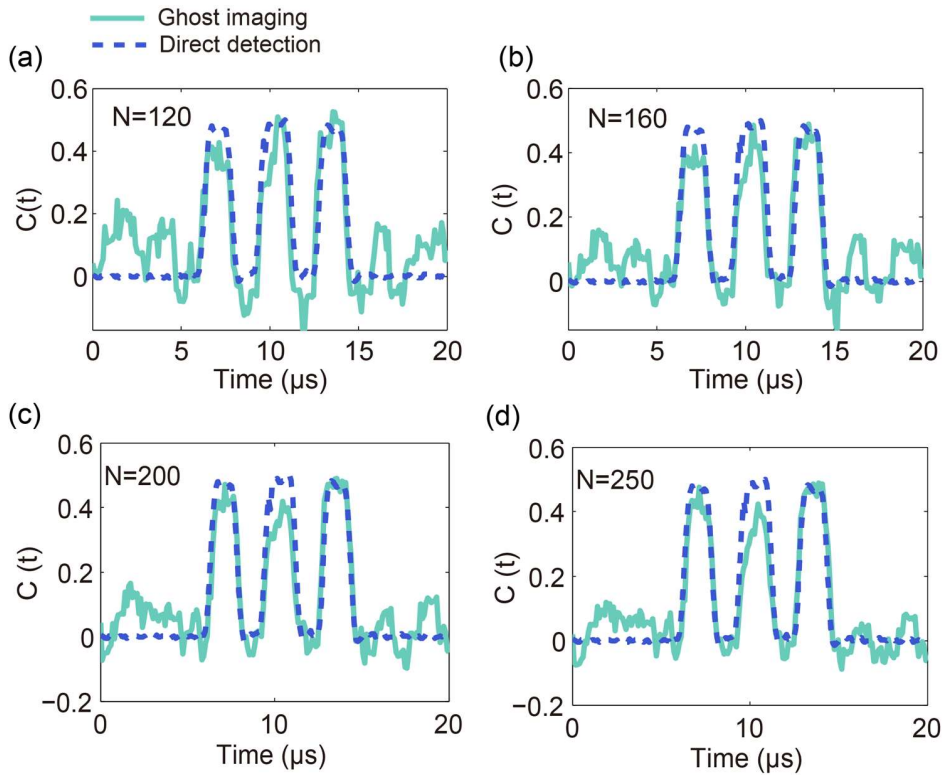

**Fig. 4** Comparisons of retrieved ghost images (cyan solid line) and direct detection result (blue dashed lines) for an increasing number of probing patterns. The speed of the temporal object was set to 625 kbps.

**Mid-infrared computational TGI with Hadamard patterns.** In order to reduce the number of probing patterns to reconstruct a given temporal object, we apply an orthogonal matrix of pre-programmed temporal patterns that modulate the signal at 1542 nm. Each probing pattern $H_k(t)$ ($k$ = 1, 2, ..., $N$) is derived from the row of a 32-order Hadamard matrix whose elements are either +1 or −1. As temporal intensity cannot be negative, we employ two distinct matrices of probing patterns $H_{ko}(t)$ where the −1 elements in $H_k(t)$ are substituted with 0 (see **Supplementary Note S2)**, and $H_{ke}(t)$ which is the opposite pattern of $H_{ko}(t)$ (i.e. the 1 elements in $H_{ko}(t)$ are replaced with 0 and the 0 elements in $H_{ko}(t)$ are replaced with 1) such that $H_k(t) = H_{ko}(t) - H_{ke}(t)$. The signal at 1542 nm is first modulated by the temporal patterns $H_{ko}(t)$ and subsequently by the temporal patterns $H_{ke}(t)$. $H_{ko}(t)$ and $H_{ke}(t)$ are transferred to the idler at 3.4 μm and, after interacting with the temporal object, the integrated intensities $B_{ko}$ and $B_{ke}$ are recorded by the slow detector for the $k^{th}$ pattern of $H_{ko}(t)$ and $H_{ke}(t)$, respectively. The temporal object is retrieved from the second-order correlation between $H_k(t)$ and $B_k$ calculated over the 32 realizations where $B_k = B_{ko} - B_{ke}$.

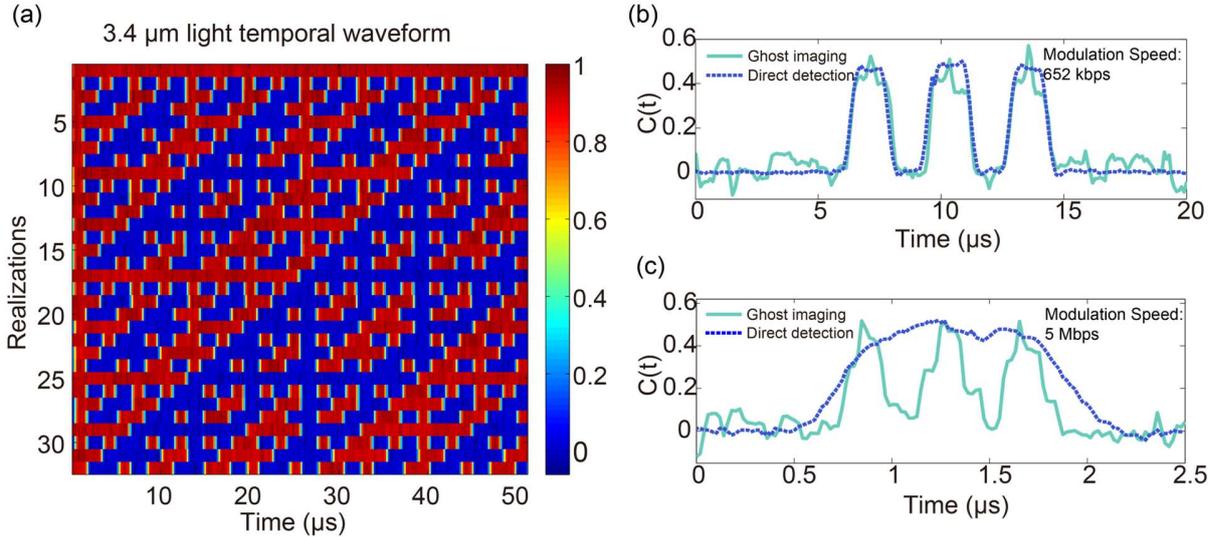

**Fig. 5** (a) Temporal waveforms measured at 3.4 μm after the PPLN crystal, confirming that the Hadamard patterns modulating the 1542 nm signal can be successfully transfer to the idler at 3.4 μm through frequency downconversion. The modulation speed of the AOM1 is set as 625 kbps. (b) Experimental result of computational TGI obtained from 32 realizations of Hadamard patterns (cyan solid line) for a 625 kbps temporal object. The Hadamard patterns are modulating the 1542 nm signal at a speed of 1.25 Mbps. (c) Experimental ghost image obtained from 32 realizations of Hadamard patterns (cyan solid line) for a 5 Mbps temporal object. The Hadamard patterns are modulating the 1542 nm signal at a speed of 10 Mbps. The blue dashed line in (b) and (c) corresponds to direct detection with a 1 MHz bandwidth mid-infrared photodetector.

The Hadamard patterns are used to modulate the temporal intensity of the 1542 nm signal and transferred to the idler at 3.4 μm through DFG. The modulation speed of the AOM1 is set as 625 kbps, and the corresponding duration of each pattern is 51.2 μs. **Figure 5(a)** shows the Hadamard patterns transferred to the temporal waveform of 3.4 μm light after the PPLN crystal directly measured with a 1 MHz bandwidth mid-infrared photodetector. Computational TGI experiments for 625 kbps and 5 Mbps temporal objects at 3.4 μm were subsequently carried out and the results obtained from the 32 probing Hadamard patterns are shown in **Figs. 5(b)** and **(c)**, respectively, along with a direct measurement of the temporal object for comparison. We can see how the computational TGI scheme with orthogonal patterns can retrieve the temporal object and this with a significant reduced number of realizations as compared to randomly selected patterns. The PSNR of the reconstructed ghost imaging obtained with 32 Hadamard patterns in **Fig. 5 (b)** is 18.91 dB, which is higher than that of the reconstructed ghost imaging obtained with 250 random patterns. Therefore, the proposed mid-infrared computational TGI could lead to significantly faster reconstruction of the temporal object in mid-infrared as compared to using a two-color detection TGI scheme utilizing random temporal intensity fluctuations.

We also experimentally investigated the influence of the mid-infrared photodetector bandwidth on the retrieval of the temporal object as shown in **Supplementary note S3**. The comparison is performed for a 5 Mbps temporal object using a mid-infrared photodetector in the test arm with 200 kHz, 500 kHz and 1 MHz bandwidth. We can see that, even for a photodetector bandwidth 25 times lower than the speed of temporal object, the temporal object is still very well retrieved. Note that it has been shown at near-infrared wavelengths that computational TGI enables to reconstruct a temporal signal with a 50 ns time scale using a detector with a bandwidth as low as 1 kHz bandwidth[22]. One can anticipate that this should also be possible with our frequency downconversion based computational TGI scheme, clearly providing new possibilities to detect ultrafast signal in mid-infrared region by using commercially available mid-infrared photodetector with MHz bandwidth.

**Temporal resolution.** The temporal resolution of our mid-infrared computational TGI is determined by the minimum temporal duration of the pre-programmed temporal patterns modulated on the 1542 nm light, which is 0.1 μs in our proof-of-concept demonstration limited by the modulation bandwidth of the AOM1. In principle, one could use lithium niobate electro-optic modulators with ultrahigh modulation bandwidth up to 40 GHz available at telecom wavelengths to increase the temporal resolution up to 25 ps. Moreover, very recently, the programmable generation of near-infrared pulses with temporal structure down to tens of femtosecond was demonstrated using a digital micromirror device and femtosecond light source[34]. Combining frequency downconversion with such ultra-fine temporal patterns in the near-infrared[34], one could further increase the temporal resolution of mid-infrared computational TGI to the sub-ps level.

**Extension to other mid-infrared wavelengths.** While in the previous section we have focused on the particular case of computational TGI at 3.4 μm, in principle the frequency downconversion scheme is flexible and can be extended to image temporal objects at other wavelengths. To experimentally demonstrate this, we tuned the pump wavelengths and corresponding phase matching conditions to generate a tunable idler from 3.2 to 4.3 μm and performed computational TGI with Hadamard patterns in this range. **Figure 6(a)** shows the spectrum of the generated idler light by tuning the pump wavelength and a fixed signal wavelength at 1542 nm. Depending on the target idler wavelength, the pump is the ytterbium-doped fiber laser with tunable range in 1040-1090 nm or a random Raman fiber laser tunable from 1110 to 1150 nm range (see **Supplementary note S1** for the lasers layout and spectral characteristics). The period and temperature of PPLN crystal are adjusted to fulfil the phase-matching conditions (see **Materials and methods** section for details). Results of computational TGI with Hadamard patterns are show in **Fig. 6(b)** for a temporal object at different mid-infrared wavelengths. One can see that, independently of the operating wavelength, the computational TGI successfully retrieve the temporal object over the full 3.2 to 4.3 μm investigated wavelength range (essentially limited by the transparency window of the PPLN crystal). To further extend the operating wavelength region beyond 5 μm, one may employ other non-oxide nonlinear crystals, such as $ZnGeP_2$ [35], orientation-patterned gallium phosphide (OP-GaP) [36], and $BaGa_4Se_7$ (BGSe)[33]. For example, DFG of 1.5 μm light modulated by the pre-programmed temporal patterns and a 2 μm tunable CW fiber laser as the pump should allow for performing computational TGI in the 6-8 μm region, while DFG with 1.3 μm tunable CW fiber laser would enable computational TGI at 10 μm [35].

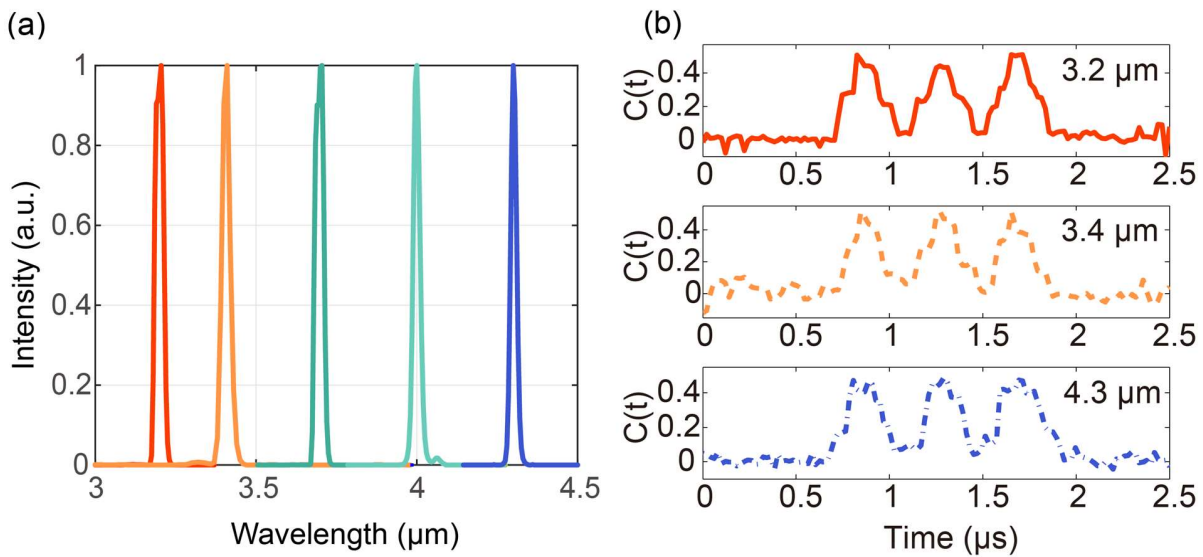

**Fig. 6.** (a). Idler spectrum generated by DFG of tunable fiber laser pump and 1542 nm signal light in the PPLN crystal. (b) Experimental temporal ghost image obtained from 32 realizations of Hadamard patterns for a 5 Mbps temporal object at selected mid-infrared wavelengths as indicated.

**Discussion**

In conclusion, we have introduced the concept frequency downconversion-TGI whereby temporal patterns modulating a CW laser signal are downconverted in a nonlinear crystal to an idler that interacts with a temporal object to be characterized. The wide availability of tunable lasers in the near-infrared[37,38] allows for flexible and versatile operation of the downconversion TGI scheme, enabling to extend TGI to wavelength regimes where there is a lack of fast detectors and modulators. Using this approach, we have experimentally demonstrated computational TGI in the wavelength range from 3.2 to 4.3 μm. We have also demonstrated computational downconversion TGI where one uses orthogonal patterns to reduce the number of distinct probing measurements. To highlight the benefits of our approach, a side-by-side comparison with existing TGI methods is provided in **Supplementary note S4.** Compared to previously reported TGI schemes, the proposed frequency downconversion-computational TGI enables imaging temporal object in the mid-infrared region by using commercially available telecom light sources and modulator and very low bandwidth mid-infrared detectors. It should also be noted that instead of using a slow mid-infrared detector that records the integrated intensity after the temporal object, upconversion detection after the temporal object could be applied to perform bucket detection at visible or near-infrared region with a slow silicon or InGaAs detector.

The temporal resolution of the downconversion computational TGI is determined by the temporal resolution of the preprogrammed temporal patterns modulated on the 1.5 μm light. Using a 1.5 μm telecom electro-optic modulator with modulation bandwidth up to 40 GHz, one can increase the speed of the temporal object up to tens of Gbps at mid-infrared, and we therefore anticipate the proposed mid-infrared computational TGI scheme to provide new possibilities to characterize advanced high-speed mid-infrared intensity modulators[39,40] and enable implementing high-speed free-space optical communications in the 3-5 μm and 8-14 μm atmospheric transmission windows[28,41] even in the presence of atmospheric turbulence[42] (see **Supplementary note S5** for a possible schematic of high-frequency transmission and secure communication in the mid-infrared region using frequency downconversion computational TGI). Furthermore, applying recently developed near-infrared programmable temporally structured pulses[34] in the frequency downconversion process, computational TGI with temporal resolution down to the sub-ps level could lead to a new generation of scan-free pump-probe ghost imaging for the study of ultrafast dynamics in the mid-infrared spectral region such as e.g. ultrafast all-optical modulation signal and semiconductor carrier lifetime measurements[29].

Finally, we emphasize that the concept of frequency downconversion ghost imaging is generic, and it can also be applied in the spatial and spectral domains, which could unlock new possibilities to realize e.g. single-pixel imaging and spectroscopy in spectral region where preprogrammed modulation is difficult to apply such as the mid-infrared and THz regions[31,32]. For example, computational spectral domain ghost imaging has been demonstrated in the near-infrared region using a programmable spectral filter[43,44] enabling to perform spectroscopy in weak light condition and strong turbulence[44]. With frequency downconversion, one can transfer the preprogrammed spectral patterns to the idler wave in mid-infrared and perform computational ghost spectroscopy in the mid-infrared region, opening up new perspectives for remote sensing in industrial, biological or security applications as many molecules display fundamental vibrational absorptions in mid-infrared spectral region[45].

**Materials and methods**

**Probing patterns generation:** 250 realizations of randomly chosen binary patterns or 32-order Hadamard matrix patterns are generated on a computer and used to drive the AWG (RIGOL DG4062, 60 MHz bandwidth, 2 channels). The output of AWG is divided into two paths, one is connected to the oscilloscope (R&S RTO2024, 2 GHz bandwidth) and another is connected to the RF driver of AOM1 (Gooch & Housego Fiber Q) to modulate the transmission of AOM1 and thus the temporal intensity of the 1542 nm signal.

**Temporal object generation:** The temporal object $S(t)$ consisting of a test bit sequence with duration $T$ is sent to the AWG that drives the transmission of AOM2 and modulates the idler intensity at the mid-infrared wavelength. AOM2 has a transmission above 95% and diffraction efficiency exceeding 80% in the 3.3-3.7 μm range, with a rise time of 140 nm/mm. In our experiments, the beam diameter of the mid-infrared light on AOM2 is designed as 1 mm. After interacting with the temporal object, the mid-infrared idler light is detected by a mid-infrared photodetector and the output of mid-infrared detector is recorded in the oscilloscope as $PD_k(t)$.

**1542 nm light temporal waveform measurement:** 10% of the modulated 1542 nm signal light after AOM1 is tapped out using a 1:9 beam splitter and detected with a near-infrared photodetector (Daheng Optics, DH-GDT-D002N, 100 MHz bandwidth).

**Direct detection:** For the direct detection measurements, AOM1 is removed, and the mid-infrared idler light after the temporal object is directly detected with the mid-infrared photodetector.

**Mid-infrared idler light characterization:** The spectrum of the mid-infrared idler light is measured with a grating-scanning monochromator (Zolix Omni-λ500i) coupled to a lock-in amplifier (SRS, SR830) and a liquid nitrogen mercury cadmium telluride detector (Judson, DMCT16-De01). The power of the idler light is measured by a power meter (Ophir, 3A, power range: 10 μW–3 W @ 0.19–19 μm).

**Data processing**: The pre-programmed temporal patterns $R_k(t)$ and the waveforms of mid-infrared detector $PD_k(t)$ measured by the oscilloscope are saved and processed offline. In the offline processing, the time-integrated output of the mid-infrared detector $B_k = \int_0^T PD_k(t')dt'$ is used to reconstruct the temporal object by computing the second-order correlation between $R_k(t)$ and $B_k$ calculated over the probing ensemble of realizations. The source code for reconstructing the temporal object is provided in **Supplementary Note S2**.

**Peak signal-to-noise ratio:** The accuracy of the retrieved temporal object can be evaluated from the peak signal-to-noise ratio (PSNR) between the directly measured and reconstructed object:

$\text{PSNR} = 10 log_{10}\left(\frac{MAX^2}{\frac{1}{K}\sum_1^K (G_i - D_i)^2}\right)$, where $G$ and $D$ are the retrieved and directly measured temporal object, respectively, and $K$ is the number of points in the temporal object sequence. $MAX$ is the peak value directly measured in the temporal sequence.

**Phase matching conditions for DFG:** The dimensions of PPLN crystal are 1 (height) × 1 (width) × 25 (length) mm³. The PPLN array contains 6 separated PPLN crystals with 6 different periods. The PPLN temperature and period applied to phase match DFG at different mid-infrared wavelengths are given in Table 1.

**Table 1** Phase-matching conditions to achieve efficient DFG in PPLN at mid-infrared wavelengths.

| Mid-infrared wavelength (μm) | Pump wavelength (μm) | PPLN period (μm) | Temperature (°C) |
|---|---|---|---|
| 3.2 | 1.042 | 30.10 | 125.8 |
| 3.4 | 1.060 | 30.10 | 148.3 |
| 3.7 | 1.088 | 30.10 | 112.5 |
| 4.08 | 1.120 | 29.52 | 100.2 |
| 4.3 | 1.135 | 29.08 | 109.3 |

**Acknowledgments**

This work was supported by National Natural Science Foundation of China (62375189, 62311530045, 62005186, 62075144), Sichuan Outstanding Youth Science and Technology Talents (2022JDJQ0031), Engineering Featured Team Fund of Sichuan University (2020SCUNG105) and the Research Council of Finland (320165, 356243 and 333949).

**Conflict of interests.** The authors declare no competing financial interests.

**Contributions.** H.W conceived the idea of this work. H.W. and G.G. designed the experiment. H.W., B.H., L.C., F.P. built the experimental setups and carried out all the experiments. H.W., B.H., Z.W., H. L. and G.G. prepared the manuscript. All the authors discussed the results and contributed the paper. Han Wu and Bo Hu contribute equally to this work.

**Data availability.** The data that support the plots and maps within this paper and other findings are available from the corresponding authors upon reasonable request.

**Supplementary Information.** All copyrights are reserved in www.nature.com